%Paper: gr-qc/9404007
%From: jbrown@unity.ncsu.edu
%Date: Wed, 6 Apr 1994 11:53:14 -0400 (EDT)

%% This is a plain TeX file.
\message{*** SIAM Plain TeX Proceedings Series macro package, version 1.0,
November 6, 1992.***}

\catcode`\@=11

\baselineskip=14truept

%%%  DIMENSIONS  %%%
 \hsize=36truepc
\vsize=55truepc
\parindent=18truept
\def\firstpar{\parindent=0pt\global\everypar{\parindent=18truept}}
\parskip=0pt

%%%  FONTS  %%%
\font\tenrm=cmr10

\font\tensmc=cmcsc10

\font\ninerm=cmr9
\font\ninebf=cmbx9
\font\nineit=cmti9
\def\ninepoint{%
   \def\rm{\ninerm}\def\bf{\ninebf}%
   \def\it{\nineit}\baselineskip=11pt\rm}%

\fontdimen13\tensy=2.6pt
\fontdimen14\tensy=2.6pt
\fontdimen15\tensy=2.6pt
\fontdimen16\tensy=1.2pt
\fontdimen17\tensy=1.2pt
\fontdimen18\tensy=1.2pt

\font\ninerm=cmr9
\font\elevenrm=cmr10 scaled\magstephalf
\font\fourteenrm=cmr10 scaled\magstep 1
\font\eighteenrm=cmr10 scaled\magstep 3
\font\twelvebf=cmbx10 scaled\magstep 1
\font\elevenbf=cmbx10 scaled\magstephalf

\def\textfont{\elevenrm}

\def\headfont{\twelvebf}
\def\smallheadfont{\elevenbf}
\def\titlefont{\eighteenrm}

\def\authorfont{\fourteenrm}
\def\rheadfont{\tenrm}
\def\abstractfont{\tenrm}

%\font\eightsmc=cmcsc8

\def\footnote#1{\baselineskip=11truept
   \edef\@sf{\spacefactor\the\spacefactor}#1\@sf
  \insert\footins\bgroup\ninepoint\hsize=36pc
  \interlinepenalty10000 \let\par=\endgraf
   \leftskip=0pt \rightskip=0pt
   \splittopskip=10pt plus 1pt minus 1pt \floatingpenalty=20000
\smallskip
\item{#1}\bgroup\baselineskip=10pt\strut
\aftergroup\@foot\let\next}
\skip\footins=12pt plus 2pt minus 4pt
\dimen\footins=36pc

%%%  CHAPTER OPENING MACROS  %%%
\def\startchapter{\topinsert\vglue54pt\endinsert}

\def\title#1\endtitle{\titlefont\centerline{#1}\vglue5pt}
   %\vskip40truept\tenrm}
\def\lasttitle#1\endlasttitle{\titlefont\centerline{#1}
   \vskip1.32truepc}
\def\author#1\endauthor{\authorfont\centerline{#1}\vglue8pt\textfont}
\def\lastauthor#1\endlastauthor{\authorfont\centerline{#1}
   \vglue2.56pc\textfont}

\def\abstract#1\endabstract{\baselineskip=12pt\leftskip=2.25pc
     \rightskip=2.25pc\abstractfont{#1}\textfont}

%%%  COUNTERS FOR HEADINGS  %%%
\newcount\headcount
\headcount=1
\newcount\seccount
\seccount=1
\newcount\subseccount
\subseccount=1
\def\secreset{\global\seccount=1}
 \def\subsecreset{\global\subseccount=1}

%%%  HEADINGS  %%%
\def\headone#1{\baselineskip=14pt\leftskip=0pt\rightskip=0pt
  \vskip17truept\parindent=0pt
{\headfont\the\headcount\hskip14truept #1}
\par\nobreak\firstpar\global\advance\headcount by 0
   \global\advance\headcount by 1\secreset\vskip2truept\textfont}

\def\headtwo#1{\advance\headcount by -1%
   \vskip17truept\parindent=0pt{\headfont\the\headcount.%
   \the\seccount\hskip14truept #1}%\enspace\ignorespaces\firstpar
   \global\advance\headcount by 1\global\advance\seccount by 1
   \global\advance\subseccount by 1\subsecreset\vskip2pt\textfont}

 \def\headthree#1{\advance\headcount by -1\advance\seccount by -1
%   \advance\subseccount by -1%
   \vskip17truept\parindent=0pt{\smallheadfont\the\headcount.%
   \the\seccount.\the\subseccount\hskip11truept #1}\hskip6pt\ignorespaces
   \firstpar\global\advance\headcount by 1\global\advance\seccount by 1
   \global\advance\subseccount by 1\textfont}

%%%  REFERENCES  %%%
\newdimen\refindent@
\newdimen\refhangindent@
\newbox\refbox@
\setbox\refbox@=\hbox{\tenrm\baselineskip=11pt [00]}%   Default 2 digits
\refindent@=\wd\refbox@

\def\resetrefindent#1{%
	\setbox\refbox@=\hbox{\tenrm\baselineskip=11pt [#1]}%
	\refindent@=\wd\refbox@}

\def\Refs{%
	\unskip\vskip1pc
	\leftline{\noindent\headfont References}%
	\penalty10000
	\vskip4pt
	\penalty10000
	\refhangindent@=\refindent@
	\global\advance\refhangindent@ by .5em
        \global\everypar{\hangindent\refhangindent@}%
	\parindent=0pt\baselineskip=12pt\tenrm}

\def\sameauthor{\leavevmode\vbox to 1ex{\vskip 0pt plus 100pt
    \hbox to 2em{\leaders\hrule\hfil}\vskip 0pt plus 300pt}}

\def\ref#1\\#2\endref{\leavevmode\hbox to \refindent@{\hfil[#1]}\enspace
#2\par}

%%%  OUTPUT  %%%
\def\rightheadline{\hfill\tensmc\rightrh\hskip2pc\tenrm\folio}
\def\leftheadline{\tenrm\folio\hskip2pc\tensmc\leftrh\hfill}

\global\footline={\hss\tenrm\folio\hss}% first page

\output{\plainoutput}
\def\plainoutput{\shipout\vbox{\makeheadline\pagebody\makefootline}%
  \advancepageno
  \ifnum\pageno>1
	\global\footline={\hfill}%
  \fi
  \ifodd\pageno
	\global\headline={\rightheadline}%
  \else
	\global\headline={\leftheadline}%
  \fi
  \ifnum\outputpenalty>-\@MM \else\dosupereject\fi}
\def\pagebody{\vbox to\vsize{\boxmaxdepth\maxdepth \pagecontents}}
\def\makeheadline{\vbox to\z@{\vskip-22.5\p@
  \line{\vbox to8.5\p@{}\rheadfont\the\headline}\vss}%
    \nointerlineskip}
\def\makefootline{\baselineskip24\p@\vskip-6\p@\line{\the\footline}}
\def\dosupereject{\ifnum\insertpenalties>\z@
   % something is being held over
  \line{}\kern-\topskip\nobreak\vfill\supereject\fi}

\def\footnoterule{\vskip11pt\kern -4\p@\hrule width 3pc \kern 3.6\p@ }
   %rule = .4 pt high

\catcode`\@=13

%%  BEGIN  %%

\def\leftrh{David Brown}
\def\rightrh{Reference Fluids as Standards of Space and Time}

\startchapter %Place this command at the beginning of the file immediately
              %after the \input command.

%%\title more title\endtitle
\lasttitle Reference Fluids as Standards of Space and Time\endlasttitle

%%\author more authors\endauthor
\lastauthor David Brown\footnote{$^*$}{Departments of Physics and
Mathematics,
North Carolina State University, Raleigh, NC 27695--8202}
\endlastauthor

\centerline{\bf Abstract}
\abstract The idea that spacetime points are to be identified by a fleet
of clock--carrying particles can be traced to the earliest days of general
relativity. Such a fleet of clocks can be described phenomenologically as
a reference fluid. One approach to the problem of time consists in
coupling the metric to a reference fluid and solving the
super--Hamiltonian
constraint for the momentum conjugate to the clock time variable. The
resolved constraint leads to a  functional Schr\"{o}dinger equation and
formally to a conserved inner product. The reference fluid that is
described phenomenologically as incoherent dust has the extraordinary
property that the true Hamiltonian density for the coupled system depends
only on the gravitational variables. The dust particles also endow space
with a privileged system of coordinates that allows the supermomentum
constraint to be solved explicitly.\endabstract

\headone{Introduction}
General relativity raises the conceptual question: How are different
spacetime points distinguished from one another? If there is a
wiggle in the gravitational field, how does one specify if it is here
or there, now or later? Presumably, the answer is that certain matter
fields, or certain features of the gravitational field, provide a physical
reference system that distinguishes spacetime events from one another.
For
example, we can fill space with a cloud of clock--carrying particles and
give each particle a unique label $Z^k$ ($k = 1$, $2$, $3$).  We then
observe that the wiggle occurs at the location of particle $Z^k =
(2,5,3)$,
for example, and specify the time of the wiggle by the reading on the
clock carried by that particle. If the detailed inner workings of the
clocks are represented by a phenomenological variable $T$, the clock
reading,
then we see that at each spacetime point $p$ the physical fields of
nature split into a particle label $Z^k(p)$, a clock reading $T(p)$, and
the remaining fields $\phi(p)$. The physical quantities of interest are
represented by $\phi(T,Z)$, the wiggle $\phi$ as a function of particle
label $Z^k$ and clock reading $T$.

The fields $T$, $Z^k$ constitute a reference fluid. As early as 1920,
Einstein discussed the concept of a reference fluid. He envisioned space
filled with a deformable body, a `reference molusc' [1]. In the 1960's
DeWitt
investigated the concept quantitatively by coupling gravity to an elastic
medium which carries a clock at each lattice site (in a continuum of
lattice sites) [2]. More recently, Isham, Kucha\v{r}, Stone, and Torre
have
developed a scheme for introducing reference fluids into general
relativity [3].
Their scheme consists of adding constraints to the gravitational action to
enforce certain coordinate conditions, then restoring coordinate
invariance
by parametrization. This process generates a matter coupling, and the
matter
source can be interpreted as a reference fluid. In particular, Kucha\v{r}
and
Torre (Ref.~[3]) studied the case of Gaussian coordinate conditions and
found
that the corresponding reference fluid has the physical properties of
a heat conducting, irrotational dust. Some technical difficulties arise in
the quantization of the resulting system, but the main shortcoming of this
approach
is that simple coordinate conditions yield reference fluids that
are physically unrealistic. This leads to the question of whether or not a
physically realistic phenomenological matter source, such as ordinary
dust, can be used effectively as a reference fluid.

\headone{Dust as a Reference Fluid}
The dust action can be viewed as a phenomenological action for a cloud
of clocks that interact only via the gravitational field. The canonical
action has the
standard form for matter that is non--derivatively coupled to the
gravitational field. The dust contributions to the super--Hamiltonian
and supermomentum constraints are [4]
$$  H_a^D = P T_{,a} + P_k {Z^k}_{,a} \ ,\qquad
    H_\perp^D = \sqrt{ P^2 + g^{ab} H_a^D H_b^D } \ .\leqno(1)$$
The canonical coordinates are the scalar fields $Z^k$ and $T$.
$\zeta^k = Z^k(x,t)$ is a set of labels that tells which dust particle
passes through the spacetime point $x$, $t$, and $\tau=T(x,t)$ is the
proper
time between a fiducial hypersurface $T=0$ and the spacetime point $x$,
$t$
as measured along a dust worldline. The momentum $P$ conjugate to $T$ is
the
Eulerian mass density of the dust, and the momentum $P_k$ conjugate to
$Z^k$ is the dust momentum density along the $T={\rm const}$ surfaces as
measured by the Eulerian observers. The dust action is invariant under a
symmetry transformation in which proper time $T$ is shifted along
the dust worldlines by an infinitesimal amount ${\vartheta}(Z)$. The
corresponding conserved charge $Q[{\vartheta}]$ expresses conservation
of mass. The dust action is also invariant under changes in the dust
worldline labels, $\delta Z^k = -{\xi}^k(Z)$. The corresponding conserved
charge $Q[{\xi}]$ is related to the dust momentum in the $T={\rm const}$
surfaces.

The canonical quantization of dust coupled to gravity has been studied in
detail in Ref.~[5]. The first step is to resolve the super--Hamiltonian
constraint $H_\perp\equiv H_\perp^D + H_\perp^G = 0$, where $H_\perp^G$
is the contribution from the gravitational field, with respect to the
momentum $P$. This yields a new system of
constraints which consists of the old supermomentum constraint
$H_a\equiv H_a^D + H_a^G = 0$ and a new Hamiltonian constraint
$$ H_{\uparrow} \equiv P  + h = 0 \ ,\qquad h \equiv  -
    \sqrt{ (H_\perp^G)^2 - g^{ab} H_a^G H_b^G}   \ .\leqno(2)$$
Note that the true Hamiltonian density $h$ is independent of the dust
variables. The constraint  $H_\uparrow$ generates a displacement of space
along the dust worldlines $Z^k = {\rm const}$, and has vanishing Poisson
brackets with itself, $\{ H_\uparrow(x),H_\uparrow(x')\} = 0$. It
follows that the Poisson brackets $\{h(x),h(x')\}$ also vanish.
Thus, by dropping the dust contribution, we find the remarkable result
that
$h=0$ and $H_a^G = 0$ together form a complete set of constraints for
vacuum general relativity that generates a true Lie algebra.
However, because of the square--root form for $h$, the system evolution
is not defined in the absence of dust.

It is convenient to make a change of variables that takes full advantage
of
the fact that the dust particles constitute a physical system of
coordinates
in space $\Sigma$. First observe that the congruence of dust worldlines
can
be viewed as an abstract space ${\cal S}$, the `dust space', whose points
$\zeta\in{\cal S}$ are the individual dust particles. Then the dust labels
$Z^k$ can be viewed collectively as the local coordinate expression of a
mapping $Z:\Sigma\to{\cal S}$. Given a space point $x\in\Sigma$,
$\zeta=Z(x)$
tells which dust particle is there. Given a dust particle $\zeta\in{\cal
S}$,
the inverse mapping $x=X(\zeta)\equiv Z^{-1}(\zeta)$ tells where that dust
particle is. The mapping $Z$ and its inverse $X$ can be used to map all
space
tensors (and tensor densities) from $\Sigma$ to the dust space ${\cal S}$.
In particular, $T(x)$ and $g_{ab}(x)$ are mapped to
${\overline T}(\zeta)$ and ${\overline g}_{k\ell}(\zeta)$ and their
canonical
conjugates are mapped to ${\overline P}(\zeta)$ and
${\overline p}^{k\ell}(\zeta)$. Under the canonical point transformation
$(\, T(x),g_{ab}(x),Z^k(x)\,) \mapsto (\, {\overline T}(\zeta),
{\overline g}_{k\ell}(\zeta),X^a(\zeta)\,)$, the new canonical momenta are
${\overline P}(\zeta)$, ${\overline p}^{k\ell}(\zeta)$, and
$P_a(\zeta) = -|\partial X(\zeta)/\partial\zeta| \, H_a(X(\zeta))$.
The constraints for the coupled system become
$${\overline H}_{\uparrow}(\zeta) \equiv {\overline P} +
   {\overline h} = 0\ ,\qquad
   {\overline H}_{k}(\zeta) \equiv - P_a {X^a}_{,k} = 0  \ ,\leqno(3)$$
where ${\overline h}(\zeta)$ is the space tensor density $h(x)$ mapped
to the dust space ${\cal S}$.
The constraint ${\overline H}_{k}(\zeta)$ is the canonical generator of
${\rm Diff}\Sigma$. As such, it does not depend on the ${\cal S}$ tensors
${\overline T}(\zeta)$, ${\overline g}_{k\ell}(\zeta)$, ${\overline
P}(\zeta)$
and ${\overline p}^{k\ell}(\zeta)$. The canonical generator of
${\rm Diff}{\cal S}$ is the conserved charge $Q[\xi]$.

Dirac quantization is implemented by imposing the constraints (3) as
operator restrictions on the quantum states. In the coordinate
representation,
the states are wave functionals $\Psi[{\overline T},{\overline g},X]$, and
${\widehat{\overline H}}_{k}(\zeta) \Psi = 0$ implies that $\Psi$
is independent of $X^a(\zeta)$. The remaining constraint takes the form of
a functional Schr\"odinger equation,
$$ i {\delta\Psi[{\overline T},{\overline g}] \over \delta {\overline
T}(\zeta) }
   = {\widehat{\overline h}}(\zeta) \Psi[{\overline T},{\overline g}] \
.\leqno(4)$$
Since ${\overline h}(\zeta)$ is given by the square root of an expression
${\overline G}(\zeta)$ that is not positive definite, ${\overline
h}(\zeta) \equiv
-\sqrt{{\overline G}(\zeta)}$, the operator ${\widehat{\overline
h}}(\zeta)$ must
be defined by spectral analysis: The commutators
$[{\widehat{\overline G}}(\zeta),{\widehat{\overline G}}(\zeta') ]$ are
assumed
to vanish, since the corresponding Poisson brackets are indeed zero. It
follows that
the operators ${\widehat{\overline G}}(\zeta)$ have common
eigenfunctionals, and
the eigenfunctionals with positive eigenvalues ${{\overline G}}(\zeta) >0$
span a
Hilbert space ${\cal H}^+$. Then
${\widehat{\overline h}}(\zeta)$ is defined as the operator in ${\cal
H}^+$ with
eigenvalues $-\sqrt{{{\overline G}}(\zeta)}$. The functional Schr\"odinger
equation
(4) yields a formally conserved inner product
$$\langle \Psi_1 | \Psi_2 \rangle = \int D{\overline g} \,
   \Psi_1^*[{\overline T},{\overline g}] \Psi_2[{\overline T},{\overline
g}]
   \leqno(5)$$
in the Hilbert space ${\cal H}^+$, where $D{\overline g}$ is a measure on
the
space of Riemannian metrics on the dust space ${\cal S}$.

The reference system provided by a phenomenological dust cloud of clocks
allows
one to go quite far in the formal construction of a quantum theory of
gravity.
Unfortunately, there is a difficulty. The allowed quantum observables
$\widehat F$ must have the property that ${\widehat F}\Psi\in{\cal H}^+$
for all
$\Psi\in{\cal H}^+$. Although the metric ${\overline g}_{k\ell}(\zeta)$ is
classically
observable, the operator ${\widehat{\overline g}}_{k\ell}(\zeta)$ (defined
as
multiplication by ${\overline g}_{k\ell}(\zeta)$ in the coordinate
representation)
is not a quantum observable. The failure of the metric tensor to qualify
as a quantum
observable undermines one of the motivations for introducing the reference
fluid
in the first place.

\Refs

\ref 1\\A.~Einstein, {\it Relativity: The Special and the General Theory},
Crown, New York, 1961.\endref

\ref 2\\B.~S. DeWitt, in {\it Gravitation: An Introduction to Current
Research},
edited by L.~Witten, Wiley, New York, 1962.\endref

\ref 3\\C.~J. Isham and K.~V. Kucha\v{r}, {\it Ann. Phys.} (NY) {\bf 164}
(1985) 288, 316; C.~G. Torre, {\it Phys. Rev.} {\bf D40} (1989) 2558;
K.~V. Kucha\v{r} and C.~G. Torre, {\it Phys. Rev.} {\bf D43} (1991) 419,
{\bf D44} (1991) 3116; C.~L. Stone and K.~V. Kucha\v{r}, {\it Class.
Quantum Grav.} {\bf 9} (1991) 757; K.~V. Kucha\v{r}, {\it Phys. Rev.}
{\bf D43} (1991) 3332, {\bf D44} (1991) 43.\endref

\ref 4\\J.~D. Brown, {\it Class. Quantum Grav.} {\bf 10} (1993)
1579.\endref

\ref 5\\J.~D. Brown and K.~V. Kucha\v{r}, submitted to {\it Phys. Rev.}
{\bf D}.\endref

\bye